# Photonic glass for high contrast structural color

Guoliang Shang,[1,*] Lukas Maiwald,[1] Hagen Renner,[1] Dirk Jalas,[1] Maksym Dosta,[2] Stefan Heinrich,[2] Alexander Petrov,[1,3] and Manfred Eich[1,4]

[1]Institute of Optical and Electronic Materials, Hamburg University of Technology, Eissendorfer Strasse 38, 21073 Hamburg, Germany

[2]Institute of Solids Process Engineering and Particle Technology, Hamburg University of Technology, Denickestrasse 15, 21073 Hamburg, Germany

[3]ITMO University, 49 Kronverkskii Ave., 197101, St. Petersburg, Russia

[4]Institute of Materials Research, Helmholtz-Zentrum Geesthacht, Max-Planck-Strasse 1, Geesthacht, D-21502, Germany

[*]Corresponding author: guoliang.shang@tuhh.de

**Abstract**

Non-iridescent structural colors based on disordered arrangement of monodisperse spherical particles, also called photonic glass, show low color saturation due to gradual transition in reflectivity. No significant improvement is usually expected from particles optimization, as the Mie resonances are broad for small dielectric particles with moderate refractive index. Moreover, the short range order of a photonic glass alone is also insufficient to cause sharp spectral features. We show here, that the combination of a well-chosen particle geometry with the short range order of a photonic glass has strong synergetic effects. We demonstrate how core-shell particles can be used to obtain a sharp transition in the reflection spectrum of photonic glass which is essential to achieve a strong color saturation. The Fourier transform required for a highly saturated color can be achieved by shifting the first zero position of the motif Fourier transform to smaller wave numbers in respect to the peak of the lattice Fourier transform. We show that this can be obtained by choosing a non-monotonous refractive index distribution from the center of the particle through the shell and into the background material. The first-order theoretical predictions are confirmed by numerical simulations.

**Introduction**

Structural color is a color based on selective light scattering and reflection from nanostructures [1-3]. The commercial pigment based color derives from light absorption by electron transitions and is dependent on the presence of a defined chemical structure, which can be altered by UV radiation during later use or high temperature processing during manufacturing [2,4]. Also, some of the pigments contain toxic materials that can be harmful in production or disposal, initiating the need for alternatives [2,5,6]. At the same time, structural colors depend on the refractive index distribution, only and thus can be produced from environmentally friendly materials such as silica, alumina, zirconia etc. and therefore bear the potential of high UV and temperature stability. Structural colors can be divided into two classes: iridescent and non-iridescent colors. An iridescent color is usually based on periodical structures with the periodical length in the order of visible light, known as photonic crystals (PhCs) [7]. The non-iridescent structural color is angle-independent which means the color impression is the same for different illumination and observation angles. Historically, many research groups focused on microstructures mimicking biological structures to achieve non-iridescent colors. For example, the feathers of many birds can exhibit bright non-iridescent structural colors [8-11]. Some birds' feathers have structures similar to random compact arrangement of spherical particles. Such disordered arrangements, also called photonic glass (PhG) in contrast to PhC, can be obtained by self-assembly of monodispersed spherical particles [12,13]. Recently PhGs have attracted a lot of attention in the field of non-iridescent structural colors [14-18]. Non-iridescent structural colors produced by amorphous structures are mostly short wavelength colors such as violet or blue [19], since the typical band of scattered wavelengths is situated at the edge of the human eye sensitivity range which leads to the impression of a pure blue or violet as even shorter scattered wavelengths do not contribute to a color mix. Longer-wavelength structural colors towards red can be produced by PhG in combination with light absorbing pigments such as carbon black or others [20-24] which take out particularly the shorter multiply scattered wavelengths which otherwise would spoil the color impression. The PhG possesses a short-range order and the Fourier transform (FT) of its permittivity distribution is a spherical shell. The bright non-iridescent structural colors directly correlate to the spherical shell shape of the FT such as observed for feathers of the male Plum-throated Cotinga [11].

However, the reported transition from the no-reflection to back-reflection regime is still rather smooth resulting in low color saturation. Some experimental effort was invested in PhG based structural colors with core-shell particles leading to marginal spectral improvement only [14,25-28], which is due to the lack of the theoretical understanding of the influence of core-shell geometry on color purity. Most of the explanations so far are based on a manipulation of Mie resonances in the particles [1,3,16,22]. At the same time, low order Mie resonances in the low-refractive-index particles are usually spectrally very wide and thus cannot lead to sharp transitions in the scattering properties. To the best of our knowledge, for the first time, we are providing a comprehensive theoretical and simulation treatment of structural colors employing photonic glasses which explains the main mechanisms of color generation and supplies clear design and synthesis rules to achieve high color saturation.

In this work, we describe the relationships between the spectrum of a non-iridescent structural color and the FT of the permittivity distribution via the first order approximation. We split the PhG structure into the disordered lattice and the repeating motif. We show that sharp transitions in the FT of the PhG structure can be obtained by tailoring the sub-structure of the motif which leads to a shift of the first zero position of the motif Fourier transform to smaller wave numbers with respect to the peak of the lattice Fourier transform, thereby pulling the complete PhG-Fourier transform to zero along a much bigger slope. For that, the refractive index of the particle shell has to be smaller or larger than those of the core and background. Thus, a non-monotonous refractive index distribution along the radial coordinate from the center of the core through the shell into the background material is required. As will be shown for particular absolute values of the refractive indices, the relative thickness of the shell has to be adjusted in order to obtain the desired reflection characteristics. Numerical simulations confirm the appearance of sharp transitions in the reflection spectra for the optimized structures.

**Results and discussion**

The light scattering properties of disordered structures with small permittivity contrast $\Delta\varepsilon(\vec{r})$ with respect to the background level can be estimated from a first-order approximation [29] which is given in the supplementary materials. Particularly helpful is the Ewald sphere construction which geometrically predicts the wavelength dependence and the directions of the scattered light [29]. Figure 1 shows the schematic representation of the Ewald sphere construction for reflection from PhG. The scatterer (Figure 1a) has the FT of $\Delta\varepsilon(\vec{r})$ in the shape of a spherical shell (Figure 1b). The thickness of the shell in reciprocal space is related to the positional order of the spheres and will be discussed later. The wave vector of the incident light is ending at the origin of the reciprocal space. The length of the wave vector $\vec{k}_{in}$ is defined by the frequency $\omega$ and speed $c$ of light and refractive index of the background material $n_b$: $k_{in} = n_b\omega/c$. The Ewald sphere has the radius of the incident wave number and is centered at the starting point of the incident wave vector. The scattering directions are defined by the scattering wave vectors $\vec{k}_s$ starting at the center of the Ewald sphere and pointing to the overlap regions between Ewald sphere and FT of $\Delta\varepsilon(\vec{r})$. When the incident light has a small wave number (long wavelength, right image in Figure 1a, b), there is no overlap between the Ewald sphere and FT of

$\Delta\varepsilon(\vec{r})$, so that the light cannot be scattered. When the wave number is increasing (intermediate wavelength, the middle image in Figure 1a, b), the Ewald sphere starts to overlap with the FT of $\Delta\varepsilon(\vec{r})$ and the incident light will be backscattered only. If we further increase the wave number (short wavelength, left image in Figure 1a, b), the light will be backscattered into a cone of light with its opening angle spreading as the wavelength is further reduced. According to that, the expected reflection of the structure is schematically shown in Figure 1c. There will be no light reflection for long wavelengths. Then, when the wavelength decreases such that the Ewald sphere overlaps with the FT of $\Delta\varepsilon(\vec{r})$, the incident light starts to be reflected. The reflected power ($P$) from a scattering volume increases proportionally to the square of the absolute value of FT of $\Delta\varepsilon(\vec{r})$ integrated over the Ewald sphere surface (ESS) (see supplementary materials):

$$P = I_0 \frac{\omega^4}{16\pi^2 c^4} \int_{ESS} \frac{|\mathcal{F}\{\Delta\varepsilon(\vec{r})\}(\vec{k})|^2}{k_s^2} g(\theta) d^2k \tag{1}$$

where $I_0$ is the intensity of the incident plane wave of light, $\theta$ is the angle between scattered $\vec{k}_s$ and input $\vec{k}_{in}$ wavevectors and for unpolarized light $g(\theta) = (1 + \cos^2\theta)/2$. Besides the dielectric strengths of the individual scatterers it is the overlap of the Ewald sphere with the square of the FT in reciprocal space which governs the reflected power. This way the light scattering can be fully analyzed from the FT of the permittivity. Please note that the FT of $\Delta\varepsilon(\vec{r})$ alone does not define the scattering directions, and $\vec{k}$ is not the scattering vector but the difference of the wave vector of the scattered and incident waves. The light-reflection transition between the no-reflection and back-reflection (shown in Figure 1c) is determined by the sharpness, i.e. the slope of the FT spectrum of $\Delta\varepsilon(\vec{r})$ at the inner boundary of the spherical shell. In other words, if we want to achieve a sharp reflection edge, the contributions of the square of the FT of $\Delta\varepsilon(\vec{r})$ inside the shell should be as little as possible and, most importantly, the transition to large values should be sharp.

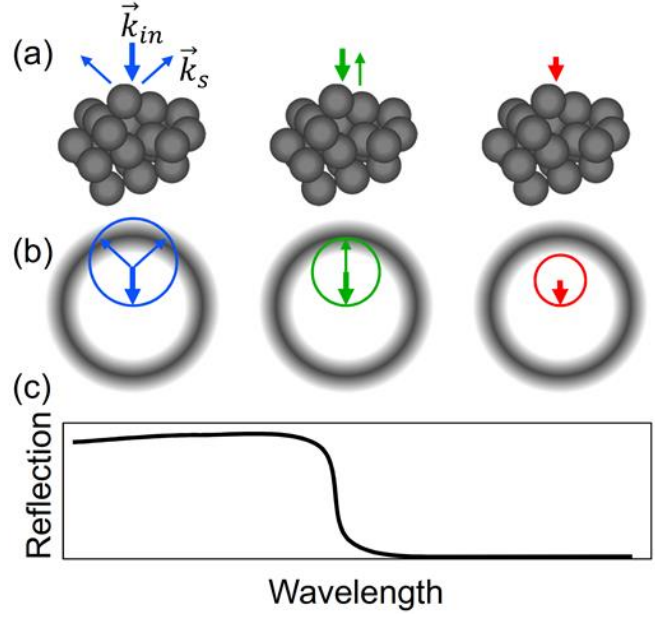

Figure 1: The schematic representation of the Ewald sphere construction for reflection from PhG. (a) Light scattered by the PhG into different wave vectors in real space. (b) The corresponding description in reciprocal space using Ewald sphere construction. The scattered wave vectors are obtained as intersection of Ewald sphere (sharp rings in red, green and blue) with the absolute squared Fourier transform of the permittivity distribution (cyan). (c) The corresponding schematic curve of the expected reflection.

In order to produce the sharp reflection edge, the square of the FT of $\Delta\varepsilon(\vec{r})$ of the PhG structure should be understood and tailored. As can be seen from Figure 2a, the PhG structure can be seen as the convolution of the disordered lattice function $l(\vec{r})$ with the motif function $m(\vec{r})$, where $l(\vec{r})$ represents the distribution of the spheres' center points in space and $m(\vec{r})$ represents the distribution of the permittivity difference in the single spherical particle. Mathematically, the FT of the whole structure is the multiplication of the lattice FT $\mathcal{F}_l(\vec{k})$ and the motif FT $\mathcal{F}_m(\vec{k})$ as shown in Figure 2b:

$$\mathcal{F}_t = \mathcal{F}\{\Delta\varepsilon(\vec{r})\} = \mathcal{F}\{l(\vec{r})\otimes m(\vec{r})\} = \mathcal{F}\{l(\vec{r})\}\cdot\mathcal{F}\{l(\vec{r})\} = \mathcal{F}_l(\vec{k})\cdot\mathcal{F}_m(\vec{k}) \qquad (2)$$

The lattice of the PhG can be characterized by the minimum distance $a$ between lattice points and the average coordination number (which is directly related to the packing density of the PhG constructed from solid spheres [30]). For the case of hard spheres as shown in Figure 2a, $a$ is equal to the particle diameter $d$. We are not interested in the particular realization of the lattice but in the average absolute square of the FTs of such lattice realizations. Such an average quantity normalized for one lattice point is also called the structure factor $S(\vec{k})$ [30,31]. $S(\vec{k})$ is a dimensionless function which approaches one for complete disorder, meaning that there is no correlation of motif positions and that the scattered intensity per particle in the matrix is equal to the intensity scattered by a single particle. In the case of incomplete disorder the correlation between particles leads to the fact that scattering into some direction is enhanced and in other directions is reduced. Thus in some directions the structure can scatter more than just the sum of intensities from single particles and $S(\vec{k})$ is larger than one. The approximate function of radial distribution $S(\vec{k})$ can be derived from solving the Ornstein-Zernike integral equation by choosing the hard sphere Percus-Yevick approximation [32-34]. Due to the spherical symmetry of the structure factor $S(\vec{k})$ in $\vec{k}$ space resulting from the isotropy of the spatial lattice, $S(\vec{k})$ is a function only of the absolute distance $|\vec{k}| = k$ from the origin of the $\vec{k}$ space and we can write $S(|\vec{k}|) = S(k)$ and assume $S(k)$ to be real:

$$S(k) = \frac{1}{1-\bar{N}C(k)} \quad (3)$$

$$\begin{cases} c(r) = 0, r > d \\ c(r) = -\xi_1 - 6\phi\xi_2\frac{r}{d} - \frac{1}{2}\phi\xi_1\frac{r^3}{d^3}, r < d \end{cases} \quad (4)$$

where $\bar{N}$ is the average number density of the spheres, i.e. the number of spheres per unit volume, $d$ is the diameter of the sphere with $a = d$ (hard spheres), $C(k)$ (the full equation of which can be found in supplementary material) is the Fourier transform of the direct correlation function $c(r)$ which represents the direct interactions between particles, and $\phi = (\pi \bar{N} d^3)/6$ is the sphere packing density. The coefficients are defined as $\xi_1 = (1 + 2\phi)^2/(1 - \phi)^4$, $\xi_2 = -(1 + \phi/2)^2/(1 - \phi)^4$. As can be seen from the equations, the structure factor depends on the sphere packing density and the minimum distance between lattice points which is equal to diameter of the sphere for the solid hard spheres. Thus, the average intensity of scattered light per motif in the lattice is proportional to the product of the structure factor $\mathcal{S}$ and the form factor $\mathcal{P}$ where the latter is the square of the motif FT normalized by the volume of the motif ($V$) and can be written as $\mathcal{P} = |\mathcal{F}_m|^2/V^2 = \mathcal{F}_m^2/V^2$ because the FT of our spherically symmetric particles is always real. Thus the average square of the FT from $N$ particles is:

$$\langle |\mathcal{F}_t|^2 \rangle = \langle \mathcal{F}_t^2 \rangle = NV^2 \mathcal{S} \cdot \mathcal{P} \tag{5}$$

In the following examples we will consider PhGs with ultimate packing density of 64% [35]. The manufacturing packing densities are slightly lower but that does not change the presented approach and conclusions.

The motif can be a homogeneous full particle, an air hole, a core-shell particle or a hollow particle, etc. The form factor amplitude of a solid sphere with permittivity contrast $\Delta\varepsilon$ has the following radial function in reciprocal space [30,36]:

$$\mathcal{F}_m(k)/V = \Delta\varepsilon \frac{3\left[\sin\left(k\frac{d}{2}\right) - \left(k\frac{d}{2}\right)\cos\left(k\frac{d}{2}\right)\right]}{\left(k\frac{d}{2}\right)^3} \tag{6}$$

where $d$ is the particle diameter. Next, we consider a particular example of a direct PhG where the motif is the silica sphere in the background air.

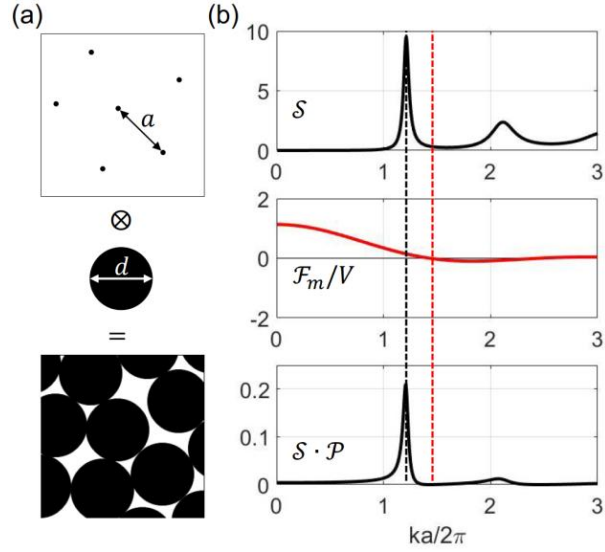

Figure 2: The permittivity of the photonic glass structure from silica spheres ($n=1.46$) and diameter $d$ in real and reciprocal space. Diameter $d$ is equal to the lattice parameter $a$ which denotes the minimum distance between sphere centers. (a) The structure can be seen as the lattice convolution with the motif, accordingly (b) the average absolute squared of the permittivity FT of the PhG structure can be written as multiplication of structure factor ($\mathcal{S}$) and form factor of the motif ($\mathcal{P}$). Instead of the form factor, the amplitude $\mathcal{F}_m(k)/V$ is presented in order to better visually identify transitions through the zero line. The form factor then results as the absolute square of the amplitude function. The horizontal axis corresponding to the radial wave number is already normalized by $2\pi/a$. The vertical red and black dashed lines indicate the zero position $k_{m0}$ of the motif and the peak position $k_{lp}$ of the lattice functions, respectively. The product ($\mathcal{S} \cdot \mathcal{P}$) of the lattice structure factor and the motif form factor is presented for the PhG assuming a packing density of 64% and a permittivity difference of $\Delta\varepsilon = 1.13$.

Figure 2b shows the result for the PhG out of homogenous solid silica spheres ($n = 1.46$) with the diameter $d = a$ and with packing density of 64% embedded in air. The product $\mathcal{S} \cdot \mathcal{P}$ is a spherically symmetric function and thus only a 1D intensity spectrum along the radial direction is shown in Figure 2b. For our considerations, the packing density is 64%, so the first peak of the lattice FT is located at about $k_{lp} = 1.21(2\pi/a)$. As can be seen, the $\mathcal{S} \cdot \mathcal{P}$ in the smaller $k$ region has relatively large intensities and first peak (located approximately at $k_{lp}$) has a smooth left edge. However, the right side of the peak has a sharper edge. This is because the first-zero point of $\mathcal{F}_m$ (at $k_{m0}$) is located at the right side of the lattice peak at $k_{lp} = 1.21(2\pi/a)$ ($k_{m0} > k_{lp}$) leading to a sharper right peak edge of the product function $\mathcal{S} \cdot \mathcal{P}$.

Having understood this mechanism, the main idea behind our work is to influence the zero position of the form factor function such that it causes a maximum slope of the product function $\mathcal{S} \cdot \mathcal{P}$ on the low-$k$ part of the spectrum, thus for the long-wavelength edge, as well as a practically empty reciporcal space towards low-$k$ numbers. These two properties eventually will lead to a reflection behavior of the photonic glass which yields a pure and highly saturated structural color. Therefore, if we move $k_{m0}$ to the left side of the peak ($k_{m0} < k_{lp}$), we could get a sharper left edge and lower intensity in the small-$k$ region. We will now follow this approach with a core-shell sphere as a motif.

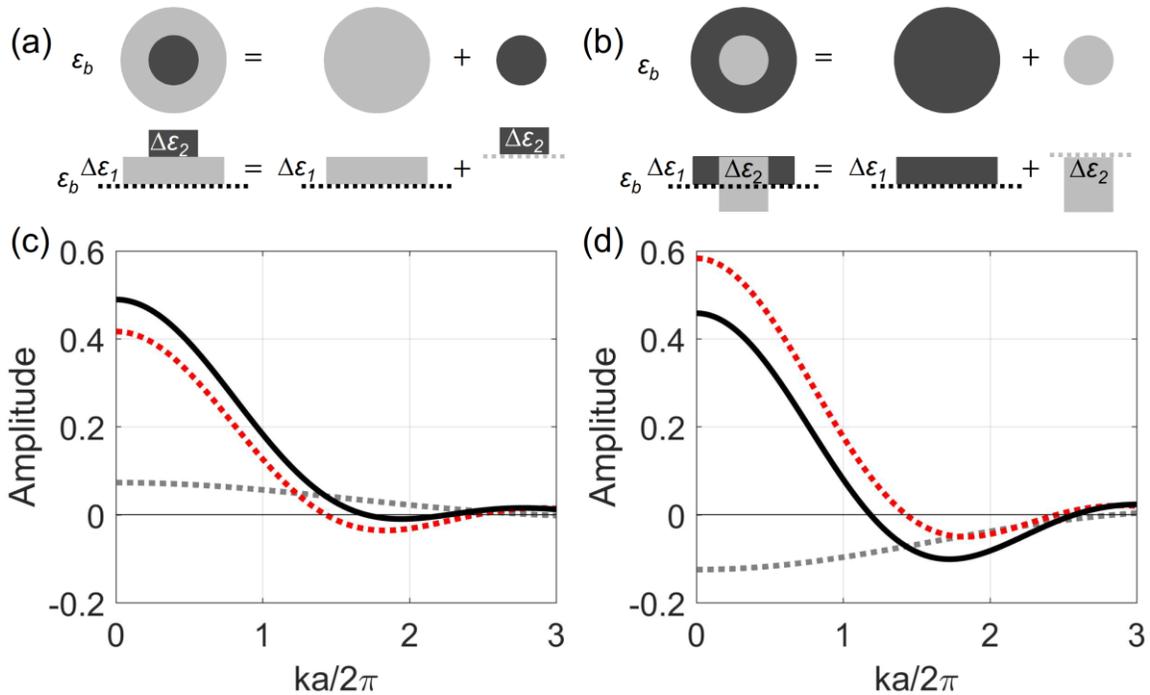

Figure 3: The normalized amplitude functions $\mathcal{F}_m/V$ for the motif of a core-shell sphere with positive contrast between shell and background and positive (a, c) and negative (b, d) refractive-index contrast between core and shell. (a, b) The core-shell sphere can be modeled as a full sphere with permittivity contrast of the shell material plus a smaller sphere with permittivity contrast between core and shell. (c, d) The corresponding amplitude functions $\mathcal{F}_m/V$ of a solid sphere consisting of the shell material, only ($\Delta\varepsilon_1$, red dot curve), of a solid sphere consisting of the core shell contrast only ($\Delta\varepsilon_2$, grey dot curve) and of the core-shell sphere (black solid curve), respectively. Subscript 1 indicates the shell and 2 indicates the core. In the case (c) implementing the core shifts the zero point to larger wave numbers. In the case (d) it allows to shift the zero point to smaller wave numbers.

When the packing method and the lattice parameter $a$ are fixed, the lattice function will not change. Accordingly, we can move the position of $k_{m0}$ by modifying sub-structure of the motif. The core-shell particle structure (Figure 3a and b) can be described as a sphere with permittivity contrast of the shell material ($\Delta\varepsilon_1 = \varepsilon_s - \varepsilon_b$) plus a smaller sphere with permittivity contrast between core and shell materials ($\Delta\varepsilon_2 = \varepsilon_c - \varepsilon_s$). So that the FT of the whole particle can be written by the following formula:

$$\mathcal{F}_m\{\Delta\varepsilon(r)\} = \mathcal{F}\{\Delta\varepsilon_1(r) + \Delta\varepsilon_2(r)\} = \mathcal{F}\{\Delta\varepsilon_1(r)\} + \mathcal{F}\{\Delta\varepsilon_2(r)\} \qquad (7)$$

First, we consider the structure with a monotonous change of the refractive index from core through shell into the background. Thus $\Delta\varepsilon_1$ and $\Delta\varepsilon_2$ can be both positive or both negative. Figure 3c shows the FT of the core-shell sphere (Figure 3a) with the corresponding refractive index of $\varepsilon_c = 2$, $\varepsilon_s = 1.5$ and $\varepsilon_b = 1$ (background is air). The ratio of the core diameter to the diameter of the whole sphere is $d_c/d = 0.5$. In this case, $\Delta\varepsilon_1 = 0.5$ and $\Delta\varepsilon_2 = 0.5$ have the same sign, $k_{m0}$ of the core-shell sphere locates at the $k$ position between the $k_{m0}$ of the solid core and of the solid shell material spheres as can be seen from Figure 3c. Thus, the $k_{m0}$ is always on the right side of the lattice peak.

Now we consider an alternative situation where the refractive index is not changing monotonously. Thus, $\Delta\varepsilon_1$ and $\Delta\varepsilon_2$ have different signs. Figure 3d shows the FT of the core-shell sphere (Figure 3b) with the corresponding refractive of $\varepsilon_c = 1$ (core is air), $\varepsilon_s = 2$ and $\varepsilon_b = 1.5$. The core-shell ratio of the diameter is the same as in Figure 3a but $\Delta\varepsilon_1 = 0.5$ and $\Delta\varepsilon_2 = -1$ have different signs. In this case the zero position $k_{m0}$ of the normalized core-shell sphere FT amplitude is located at a $k$ position smaller than $k_{m0}$ of the core or shell material as can be seen from Figure 3d. Thus, by a proper choice of refractive index contrast and shell thickness the motif zero point can be moved to the left side of the lattice peak. We have derived an equation to determine shell thickness at given refractive index contrast in the supplementary materials.

We now consider a particular core-shell particle system to implement a sharp transition in the reflection spectrum. We first optimize the FT of the structure and later compare finite integration technique (FIT) simulations for PhGs with and without optimization. Figure 4 shows $\mathcal{S}$, $\mathcal{F}_m/V$ and $\mathcal{S} \cdot \mathcal{P}$ for the hollow sphere PhG of zirconia particles with air-cores and with a background of air. We consider this particular example as such particles can be readily synthesized and such PhG-structures can be obtained by co-assembly [37]. More examples of combinations of different materials for core and shell are presented in the supplementary materials. The refractive indices are $n_c = 1$, $n_s = 2.12$, and $n_b = 1$ for the core-shell sphere. The particle diameter and minimum lattice distance are $d = a = 221.8$ nm to obtain transition for blue color, the considered packing density of the spheres is assumed to be the theoretical limit for PhG of 64%. As can be seen from Figure 4, the $k_{m0}$ can be moved from the right side to the left side of $k_{lp}$ when we increase $d_c/d$ from 0.4 to 0.9. For $d_c = 86.2$ nm shown in Figure 4a, the zero point of $\mathcal{F}_m$ is located on the right side of the lattice peak ($k_{m0} > k_{lp}$). In this case, the left side edge of the product $\mathcal{S} \cdot \mathcal{P}$ is very smooth and there is a substantial contribution at smaller $k$ values. When we increase the core diameter to 135.5 nm, the zero position shifts to a smaller $k$ value, and just overlaps with the peak ($k_{m0} \approx k_{lp}$). As can be seen from Figure 4b, the first peak of $\mathcal{S} \cdot \mathcal{P}$ almost vanishes. When we further increase the diameter of the core to 198 nm, $k_{m0}$ is moved to the left side of the lattice peak ($k_{m0} < k_{lp}$). As can be seen from Figure 4c, the left side edge of the first peak of $\mathcal{S} \cdot \mathcal{P}$ is very steep, and now the small-$k$ region has very low intensity. This kind of motif should lead to sharp reflection transition with low scattering for larger wavelengths.

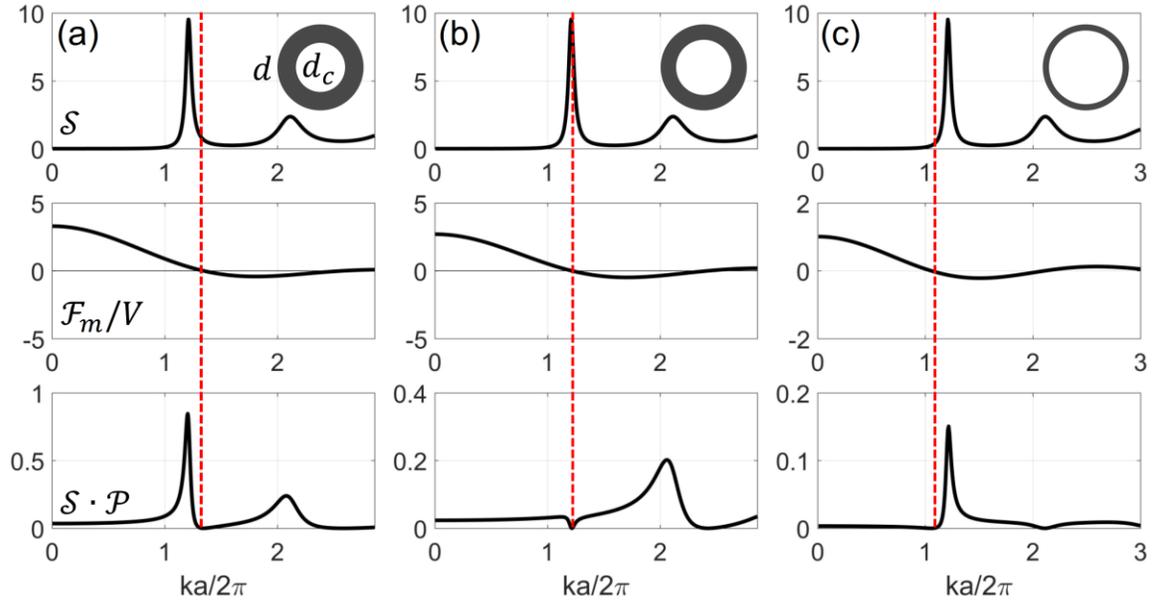

Figure 4: Lattice structure factor $\mathcal{S}$ (top), normalized motif Fourier transform amplitude $\mathcal{F}_m/V$ (middle) and product of lattice structure factor $\mathcal{S}$ and motif form factor $\mathcal{P} = \mathcal{F}_m^2/V^2$ (bottom) as functions of wavenumber $k$ for a PhG made of hollow core zirconia particles with sphere diameter $d = a = 221.8$ nm and core diameter $d_c$ of (a) 86.2, (b) 135.5 and (c) 198 nm and with background of air. The packing density is set to 64%. Insets are illustrations of the motif structure. Vertical red dashed lines indicate the zero position $k_{m0}$ of the normalized motif Fourier transform amplitude.

The main finding of the presented investigation of the influence of the varying core diameter in relation to the sphere diameter in core-shell particles with non-monotonous refractive index distribution is the fact that the zero position of the motif FT can be positioned anywhere from the right to the left side of the peak of the lattice structure factor function. Therefore, we obtain a design degree of freedom which allows us to manipulate the overall FT of the PhG, thus the quality of its structural color properties, by mere changing of the particle geometry.

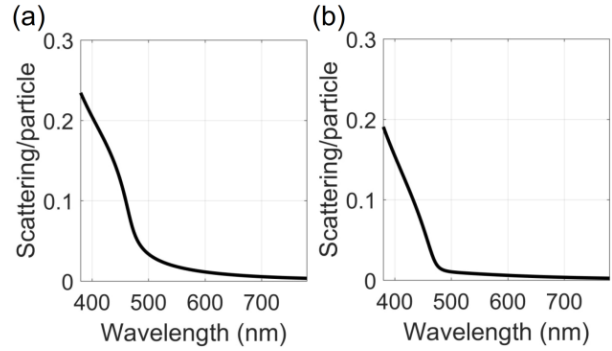

Figure 5: (a) Scattering efficiency per particle evaluated for silica sphere PhG with the FT shown by Figure 2b, assuming a diameter $d = 215.6$ nm and (b) for a hollow zirconia sphere PhG with the FT shown by Figure 4c, with $d_c = 198$ nm and $d = 221.8$ nm.

To predict the reflection curves we calculated the effective scattering cross section per single particle in the PhG:

$$\sigma = \frac{\omega^4}{16\pi^2 c^4} \int_{ESS} \frac{V^2 S(k) \cdot P(k)}{k^2} g(\theta) d^2 k \qquad (8)$$

It can be derived as the effective power scattered per single particle divided by the incident light intensity using equations (1) and (5). Figure 5 shows the scattering cross section per particle normalized by the particle geometrical cross section which represents the particle scattering efficiency compared for two PhGs: the homogenous solid silica particle PhG (with the FT shown by Figure 2b) is shown by Figure 5a and the core-shell particle PhG (with the FT shown by

Figure 4c) is shown by Figure 5b, respectively. As can be seen from Figure 5, the core-shell particle PhG shows a sharper reflection transition than the homogenous particle PhG. The sharper the left edge of the first peak of the permittivity FT-function, the sharper the right edge in the wavelength dependent reflection spectrum, hence the better the expected saturation of the structural color. It should be noted that the scattering wavevector $k_s = n_m \omega / c$ ($n_b = 1$) is used as the radius of the Ewald sphere, where $n_m$ is the mean refractive index of the PhG structure. Different sphere diameters are chosen to compensate for the different mean refractive index and by that match the transition wavelengths.

The applicability of the first-order approximation is limited to a small refractive index contrast and thin PhG films, e.g. for film thicknesses which do not substantially change the power of the light propagating through the film. For the zirconia based structures considered here, the presented approach thus can be considered only as an approximate solution. Therefore, a very important question we needed to answer was that on the predictive power of our first-order approach in view of real structures with substantial refractive index contrasts. To address this question we performed brute-force 3D FIT simulations on randomly packed PhG assemblies of core-shell spheres.

We have numerically simulated the PhGs presented in Figure 5. The structure realizations of both randomly packed PhGs of homogenous silica and of hollow zirconia spheres with 64% packing density were obtained by the packing generator MUSEN [38,39]. The periodic boundary condition was used in the packing algorithm to avoid packing density variations at the edges of the packed volume. Under normal incidence, the light reflectance ($R$) of the PhG was simulated by using the finite integration technique simulation with CST Microwave Studio [40]. For the homogeneous solid silica particle PhG, the size of the simulated structure is 2.5×2.5×12.3 µm$^3$ (number of particles is 9122). The corresponding analytical permittivity FT of the structure is shown in Figure 2b. Here, we assume a sphere diameter of 215.6 nm. For the hollow zirconia sphere PhG, the size of the simulated structure is also 1.6×1.6×7.9 µm³ (number of particles is 2961). The simulated volume size here is adjusted to make the maximal reflection comparable to that of solid sphere PhGs. The corresponding analytical permittivity FT of the structure composed of particles with 221.8 nm outer diameter and air cores of 198 nm diameter is shown in Figure 4c.

The PhGs are then excited by plane wave incident vertically from air. The lateral sides of the simulation volume are mirrors such that light can exit the simulation volume only through the open boundaries at the top and the bottom. The bottom of the PhG is adjusted to the homogeneous substrate material with refractive index equal to the average refractive index of PhG in order to minimize reflections from this boundary. The homogeneous substrate material is then terminated by an open boundary condition. The reflected power is calculated as Poynting vector integration over the upper boundary.

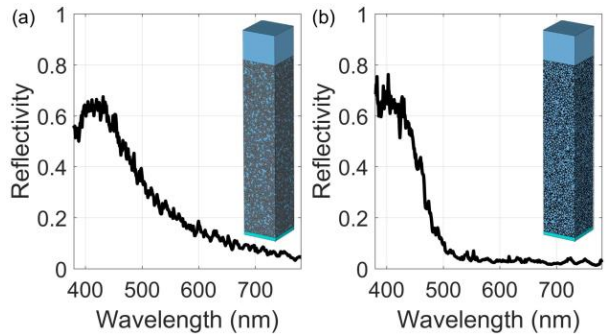

Figure 6: The comparison of simulated light reflection spectra for (a) solid silica-sphere direct PhG with the sphere size of $d = 215.6$ nm, $a = 215.6$ nm and layer thickness of 12.3 µm and (b) hollow zirconia sphere PhG with sphere size of $d = 221.8$ nm, $d_c = 198$ nm, $a = 221.8$ nm and layer thickness of 7.9 µm. Insets are the corresponding simulated structures of silica sphere direct PhG and core-shell sphere PhG.

The light reflection spectra for (a) the solid silica sphere PhG and (b) the hollow zirconia sphere PhG are shown in Figure 6. These light reflection spectra show the same trend as described in Figure 1c. In the longer wavelength region, there is nearly no light reflection. When the wavelength of incident light decreases, the Ewald sphere starts to overlap with permittivity FT of the structure, so the incident light starts to be reflected. The remarkable detail is that the light reflection spectrum of the hollow sphere PhG shows a much sharper transition compared to that of the homogenous sphere PhG. There is a good correspondence at small reflection values between the scattering efficiency per particle and the simulated reflectivity functions of the PhG-films (compare Figure 5 and 6). The enhancing effect of the presented sharp transition in reflectivity on blue color appearance is discussed in the supplementary material. For reflectivity larger than 50% the first-order approximation is not applicable as the incident wave becomes strongly depleted and the assumption of the same incident intensity on each particle is not justified anymore.

However, we are interested in analyzing the slope of the transition from very low reflection at long wavelengths to high reflection at shorter wavelengths as this slope is responsible for the saturation of the structural color. As it turned out this slope is very well predicted by the first-order approximation even considering large refractive indices, such as that of zirconia. We can therefore state that the first-order Ewald sphere approach is a simple and very well-functioning technique for predicting the quality of structural colors.

**Conclusion**

In conclusion, the selective reflectivity of the PhG structure is related to the spherical shell-shaped Fourier transform of its permittivity distribution. To explain the connection between the two properties we employ the Ewald sphere construction resulting from first-order approximation. For sharp spectral selectivity the radial distribution of permittivity Fourier transform should obtain a peak with a sharp edge at lower wave numbers. We have shown that core-shell particles can be used with non-monotonous refractive-index distribution to achieve this property. Namely, the permittivity difference $\Delta\varepsilon_1$ between shell and background should have opposite sign compared to the permittivity difference $\Delta\varepsilon_2$ between shell and core. That is, the Fourier transform of the photonic glass can be modified by changing the sub-structure of the PhG motif. In the optimal situation the zero point of the motif Fourier transform is positioned just at the small wave number edge of the peak corresponding to the Fourier transform of the PhG lattice. Numerical simulations using the finite-integration time-domain simulation confirm that the structure with optimized motif has sharp reflection transition. A particular example of hollow sphere was presented. But much more combinations are possible when two materials are combined in the core-shell sphere or background porosity of PhG is filled with a third material. The proposed innovative approach paves the road for novel structural colors with high color saturation.

40. CST, Darmstadt, Germany, Microwave Studio software. www.cst.com

**Acknowledgements**

The authors gratefully acknowledge financial support from the German Research Foundation (DFG) via SFB 986 "Tailor-Made Multi-Scale Materials Systems: M³", projects C2, A3 and priority program SPP 1839 "Tailored Disorder". The authors also acknowledge the support from CST, Darmstadt, Germany, with their Microwave Studio software.

**Author contributions**

G.S. performed the simulations, A.P. and M.E. supervised the project, G.S., L.M., H.R., D.J., M.D., S.H., A.P. and M.E. analysed the results and wrote the paper.

**Competing financial interests**

The authors declare no competing financial interests.

# Photonic glass for high contrast structural color

Guoliang Shang,[1,*] Lukas Maiwald,[1] Hagen Renner,[1] Dirk Jalas,[1] Maksym Dosta,[2] Stefan Heinrich,[2] Alexander Petrov,[1,3] and Manfred Eich[1,4]

[1]Institute of Optical and Electronic Materials, Hamburg University of Technology, Eissendorfer Strasse 38, 21073 Hamburg, Germany

[2]Institute of Solids Process Engineering and Particle Technology, Hamburg University of Technology, Denickestrasse 15, 21073 Hamburg, Germany

[3]ITMO University, 49 Kronverkskii Ave., 197101, St. Petersburg, Russia

[4]Institute of Materials Research, Helmholtz-Zentrum Geesthacht, Max-Planck-Strasse 1, Geesthacht, D-21502, Germany

[*]Corresponding author: guoliang.shang@tuhh.de

## 1. First order approximation

Here we derive the scattered-power equation using the first-order approximation in the Fraunhofer limit. The plane wave with electric field $E_{in}$ and frequency $\omega$ incident on the small volume of scattering medium $dV$, much smaller than the wavelength of light, with small permittivity contrast $\Delta\varepsilon$ generates first-order excess polarization $\varepsilon_0 \Delta\varepsilon E_{in}$, which can be considered as a dipole emitting new spherical wave with electric field amplitude in far field [1]:

$$dE = \frac{dV \Delta\varepsilon E_{in}}{4\pi} \frac{\omega^2}{c^2} \sin\gamma \frac{\exp(-i\vec{k}_s \vec{r})}{|\vec{r}|} \tag{S1}$$

where $\vec{k}_s$ is the vector with the length equal to the wavenumber of the scattered light inside the scattering medium and the direction along the scattering direction $\vec{r}$, so that $\vec{k}_s \vec{r} = |\vec{k}_s||\vec{r}|$. The angle $\gamma$ is the angle between $\vec{k}_s$ and the polarization of the incident electric field $\vec{E}_{in}$. We omit the temporal dependence $\exp(i\omega t)$ in the equations. The contribution from all excited dipoles at positions $\vec{r}'$ sums up to the total scattered field:

$$E(\vec{r}) = \frac{\omega^2}{c^2} \int \frac{\Delta\varepsilon(\vec{r}') \hat{E}_{in} \exp(-i\vec{k}_{in}\vec{r}')}{4\pi} \sin\gamma' \frac{\exp[-i\vec{k}'_s(\vec{r}-\vec{r}')]}{|\vec{r}-\vec{r}'|} d\vec{r}'^3 \tag{S2}$$

where the local polarization is now modulated with the phase $\vec{k}_{in}\vec{r}'$ of the incident wave $\hat{E}_{in}\exp(-i\vec{k}_{in}\vec{r}')$, the $\vec{k}'_s$ vector has the length equal to the wavenumber of the scattered light inside the scattering medium and the direction along the scattering direction $\vec{r} - \vec{r}'$ and $\gamma'$ is the angle between $\vec{k}'_s$ and the polarization of the incident electric field $\vec{E}_{in}$. The origin of the vector $\vec{r}'$ is defined in the vicinity of the scattering volume. In the far field several simplifications can be made, such as $|\vec{r} - \vec{r}'| \approx r$, $\vec{k}'_s \approx \vec{k}_s$ where $\vec{k}_s || \vec{r}$, and $\gamma' \approx \gamma$, thus obtaining:

$$E(\vec{r}) = \frac{\hat{E}_{in}}{4\pi} \frac{\omega^2}{c^2} \sin\gamma \frac{\exp(-i\vec{k}_s \vec{r})}{r} \int \Delta\varepsilon(\vec{r}') \exp[i(\vec{k}_s - \vec{k}_{in})\vec{r}')] d\vec{r}'^3 \tag{S3}$$

The integral now has the form of a Fourier transform. Having the total electric field and incident intensity $I_0 = nc\varepsilon_0 \hat{E}_{in}^2/2$, the intensity emitted in the far field radially into the direction $\vec{k}_s || \vec{r}$ can be written as

$$I(\vec{r}) = I_0 \frac{\omega^4}{16\pi^2 c^4} \sin^2\gamma \frac{1}{r^2} \left| \mathcal{F}\{\Delta\varepsilon(\vec{r}')\}(\vec{k}_s - \vec{k}_{in}) \right|^2 \tag{S4}$$

where $\mathcal{F}\{\Delta\varepsilon(\vec{r}')\}(\vec{k}_s - \vec{k}_{in}) = \int \Delta\varepsilon(\vec{r}') \exp[i(\vec{k}_s - \vec{k}_{in})\vec{r}')] d\vec{r}'^3$ is the three dimensional Fourier transform of the spatial function of the permittivity contrast $\Delta\varepsilon(\vec{r}')$. Thus the scattered power can be obtained by the integral of the intensity on the spherical surface (SS) in the far field:

$$P = I_0 \frac{\omega^4}{16\pi^2 c^4} \int_{SS} \sin^2\gamma \frac{\left| \mathcal{F}\{\Delta\varepsilon(\vec{r})\}(\vec{k}_s - \vec{k}_{in}) \right|^2}{r^2} d^2 r_\perp \tag{S5}$$

Due to the fact that $\vec{k}_s || \vec{r}$ and thus $\vec{r}/|\vec{r}| = \vec{k}_s/|\vec{k}_s|$, the integration over solid angle is the same in real and reciprocal space $d\Omega = (d^2 r_\perp)/r^2 = (d^2 k_\perp)/k_s^2$. Thus the power can be also calculated by integration in reciprocal space over the surface of a sphere with radius $k_s$. Note that since $\vec{r}/|\vec{r}| = \vec{k}_s/|\vec{k}_s|$ the integration is over the solid angle for vectors $\vec{k}_s$, thus in k-space the integral is on the spherical surface defined by the end points of vector $\vec{k}_s$. We prefer to integrate in the k-space over vector $\vec{k} = \vec{k}_s - \vec{k}_{in}$. In this case the integration is done on the spherical surface of a sphere with radius $k_s$ and with a center shifted from the origin by vector $-\vec{k}_{in}$. This sphere can also be identified as the Ewald sphere [1]. For the unpolarized light the $\sin^2\gamma$ can be substituted by $g(\theta) = (1 + \cos^2\theta)/2$, where $\theta$ is the angle between scattered $\vec{k}_s$ and input $\vec{k}_{in}$ wavevectors. This factor can be derived from the fact that unpolarized light has 50% of polarization always with angle $\gamma = 90°$ to the direction of scattering and another 50% with angle $\gamma = 90° - \theta$. The total scattered power is then:

$$P = I_0 \frac{\omega^4}{16\pi^2 c^4} \int_{ESS} \frac{\left| \mathcal{F}\{\Delta\varepsilon(\vec{r})\}(\vec{k}) \right|^2}{k_s^2} g(\theta) d^2 k \tag{S6}$$

integrated over the Ewald sphere surface (ESS).

## 2. Fourier transform $C(k)$ of the direct correlation function $c(r)$

The Fourier transform $C(k)$ of the direct correlation function $c(r)$ is given by [2]:

$$C(k) = \frac{24\phi}{\bar{N}}\left[\frac{\xi_1+6\phi\xi_2+\frac{1}{2}\phi\xi_1}{(dk)^2}\cos(dk) - \frac{\xi_1+12\phi\xi_2+2\phi\xi_1}{(dk)^3}\sin(dk) - \frac{2(6\phi\xi_2+3\phi\xi_1)}{(dk)^4}\cos(dk) + \frac{12\phi\xi_2}{(dk)^4} + \right.$$

$$\left.\frac{12\phi\xi_1}{(dk)^5}\sin(dk) + \frac{12\phi\xi_1}{(dk)^6}(\cos(dk)-1)\right] \tag{S7}$$

where $\bar{N}$ is the average number density of the spheres, $\phi = (\pi\bar{N}d^3)/6$ is the sphere packing density. The coefficients are defined as $\xi_1 = (1+2\phi)^2/(1-\phi)^4$, $\xi_2 = -(1+\phi/2)^2/(1-\phi)^4$.

## 3. Examples of core-shell particle structures

Here we consider several examples obtained by combination of four materials: air, silica, alumina and zirconia with refractive indexes of 1, 1.46, 1.68 and 2.12, respectively.

Figure S1a shows $\mathcal{F}_m/V$ of homogeneous zirconia sphere ($n = 2.12$) embedded in silica matrix ($n = 1.46$) with different core size. This can be implemented as zirconia spheres (diameter is $d_c$) with silica shells (the diameter $d$ of the shell is equal to $a$) and silica background infiltration. So, the sphere size is $0 < d_c/d \leq 1$. As can be seen from Figure S1a, $k_{m0}$ is always located at the right side of the lattice peak at $k_{lp}$ ($k_{m0} > k_{lp}$) when $d_c/d$ increases from 0.14 to 1.00. Similarly, the structure has the relationships of $k_{m0} > k_{lp}$ for the inverse structure of silica spheres in zirconia matrix which has a negative $\Delta\varepsilon_1$.

Figure S1b shows $\mathcal{F}_m/V$ of zirconia@silica core-shell sphere with background of air with $n_c = 2.12$, $n_s = 1.46$ and $n_s = 1$. As can be seen from Figure 4b, the $k_{m0}$ is again located at the right side of the lattice peak at $k_{lp}$ ($k_{m0} > k_{lp}$) for all $d_c/d$. For this kind of core-shell sphere, both $\Delta\varepsilon_1 = 1.13$ and $\Delta\varepsilon_2 = 2.36$ have positive values.

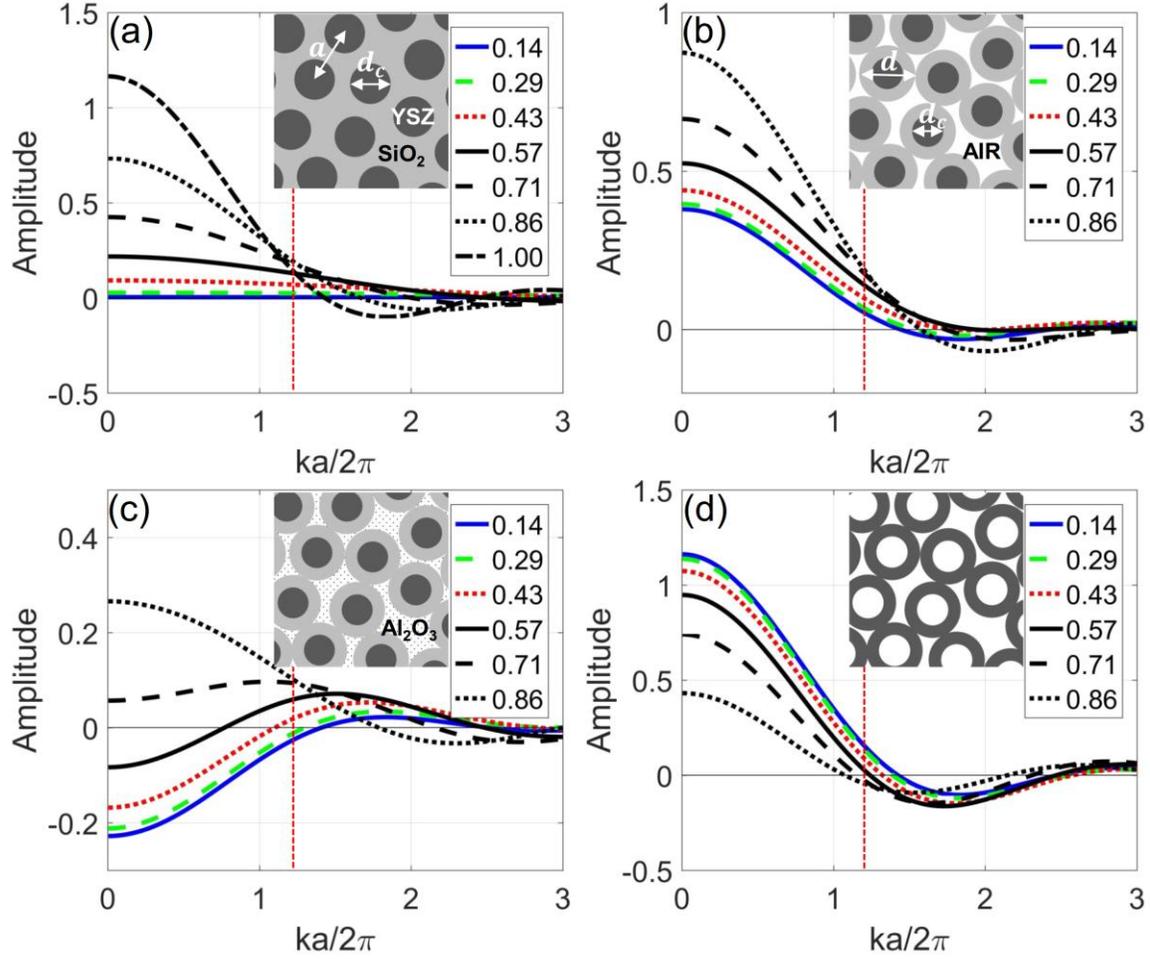

Figure S1: The Fourier transform of the motif divided by sphere volume $F_m/V$ with different structures. (a) Zirconia spheres embedded in the silica matrix. The structure also can be described as $n_c = 2.12$, $n_s = 1.46$, and $n_b = 1.46$; (b) The zirconia@silica core-shell sphere with background of air. $n_c = 2.12$, $n_s = 1.46$, and $n_b = 1$. In this case, $n_c > n_s > n_b$; (c) The zirconia@silica core-shell sphere with background of alumina ($n_b = 1.68$). In this case, $n_c > n_s < n_b$; (d) The air@zirconia core-shell sphere with background of air. Herein, $n_s > n_c = n_b$. The legends in sub-images are values of $d_c/d$. The red vertical dash line indicates the peak position $k_{lp} = 1.21(2\pi/a)$ with packing density of the spheres of 64%. The insets are the corresponding PhG illustrations.

Figure S1c shows $\mathcal{F}_m/V$ of the zirconia@silica core-shell sphere with background of alumina ($n_b = 1.68$). In this case, $n_c > n_s < n_b$ with $n_c = 2.12$, $n_s = 1.46$ and $n_b = 1.68$. As can be seen from Figure S1c, $k_{m0}$ can be shifted to the left side of the lattice peak if $d_c/d$ is in the range from 0.43 to approximately 0.65. For this kind of core-shell particles, $\Delta\varepsilon_1 = -0.69$ and $\Delta\varepsilon_2 = 2.36$ have different signs, which can be used to shift the $k_{m0}$ to smaller $k$ positions by modifying the core size. Similarly, if $\Delta\varepsilon_1 < 0$ and $\Delta\varepsilon_2 > 0$, we can also achieve the result of $k_{m0} > k_{lp}$, for example with core-shell particles with low index core, high-index shell and background material with intermediate index.

Also, there is a special case that $\Delta\varepsilon_1$ and $\Delta\varepsilon_2$ have different signs and the core material is the same as the particle surrounding material which means $|\Delta\varepsilon_1| = |\Delta\varepsilon_2|$. Example of this structure is PhG out of hollow particles discussed in the main text. Figure S1d shows $\mathcal{F}_m/V$ of the air@zirconia core-shell sphere with background of air so $n_c = 1$, $n_s = 1.46$ and $n_b = 1$. $k_{m0}$ can be moved to the left side when $d_c$ is larger than approx. 0.7.

## 4. Condition for the zero point

For the defined relationship between $\Delta\varepsilon_1$ and $\Delta\varepsilon_2$ of the core-shell particle, based on eq. 6, the zero position of the motif FT can be calculated by $\mathcal{F}_{\Delta\varepsilon_1}(k) + \mathcal{F}_{\Delta\varepsilon_2}(k) = 0$, so we obtain the following equation:

$$-\frac{\Delta\varepsilon_1}{\Delta\varepsilon_2} = \frac{\sin(k\frac{d_c}{2}) - (k\frac{d_c}{2})\cos(k\frac{d_c}{2})}{\sin(k\frac{d}{2}) - (k\frac{d}{2})\cos(k\frac{d}{2})} \tag{S8}$$

The required situation is the zero point of motif FT located at the left side of the peak of the lattice FT, that is $k_{m0} < k_{lp}$, where $k_{lp} = 1.21(2\pi/a)$ is the normalized first peak position of lattice FT as shown in Figure 2b. Because $d_c \leq d$, the right side of eq. S8 always has positive value when $kd/2 \in (0, 1.21\pi)$, that means only if $\Delta\varepsilon_1$ and $\Delta\varepsilon_2$ have different sign, the requirement of $k_{m0} < k_{lp}$ can be fulfilled. Otherwise, the zero point of the motif FT is always located at the right side of the peak of the lattice FT. For the core-shell particles ($d_c < d$), when we know the overall particle size $d$ and the refractive index of the material, the desired $k_{m0}$ can be achieved by changing $d_c$ based on eq. S8.

## 5. Color appearance

Figure S2 shows the reflection spectra of (a) the solid silica and (b) the hollow zirconia particle PhGs as well as the color matching functions of the CIE 1931 XYZ color space. Figure S2c shows the resulting points in the chromaticity diagram. In such a diagram a fully saturated color originating from a single wavelength can be found on the outer perimeter. A completely unsaturated color like grey or white is located in the so called white point in the center the diagram. For the solid spheres we are at at x=0.24 and y=0.25. The core-shell particles with their sharp reflection edge are further away from the white point yielding x=0.187 and y=0.092 and thus a more strongly saturated blue.

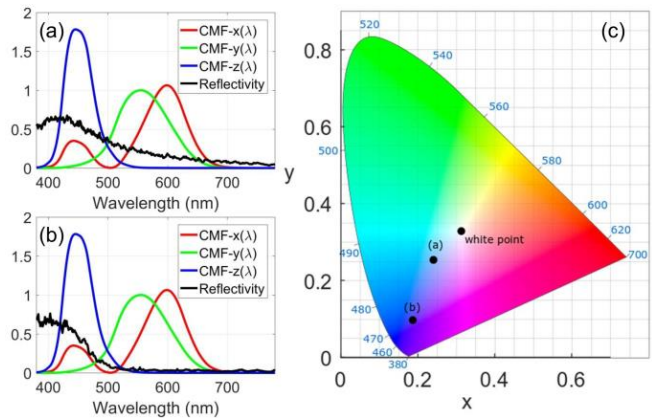

Figure S2: Reflection spectra of (a) the solid silica and (b) the hollow zirconia sphere PhGs and the matching functions (CMF) for the CIE 1931 color space. (c) Chromaticity diagram: The resulting positions of the two spectra in (a) and (b) are shown. The diagram further shows the white point and the positions of the pure colors on the perimeter indicated by the corresponding wavelengths in nm.